\documentclass[aps,prl,reprint,superscriptaddress]{revtex4-2}
\usepackage{amsmath,amssymb}
\usepackage{graphicx}
\usepackage{newtxtext}
\usepackage{newtxmath}
\usepackage{verbatim}
\usepackage[x11names]{xcolor}
\usepackage[unicode,colorlinks,citecolor=Blue3,linkcolor=Blue3,bookmarks=true]{hyperref}
\usepackage[normalem]{ulem}
\usepackage{multirow}
\usepackage{float}

\newcommand{\eg}{{\it e.g.\,}}
\newcommand{\hc}{\mathrm{h.c.}}
\newcommand{\const}{\mathop{\rm const}\nolimits}
\newcommand{\fulld}[2]{\dfrac{d#1}{d#2}}

\newcommand{\ket}[1]{|{#1}\rangle}
\newcommand{\bracket}[2]{\langle{#1}|{#2}\rangle}

\newcommand\ch[1][]{\cosh #1}
\newcommand\sh[1][]{\sinh #1}

\begin{document}


\title{A new class of pure non-Gaussian quantum states}

\author{V.~L.~Gorshenin}
\affiliation{Russian Quantum Center, Skolkovo IC, Bolshoy Bulvar 30, bld.\ 1, Moscow, 121205, Russia}
\affiliation{Moscow Institute of Physics and Technology, 141700 Dolgoprudny, Russia}

\author{B.~N.~Nougmanov}
\affiliation{Russian Quantum Center, Skolkovo IC, Bolshoy Bulvar 30, bld.\ 1, Moscow, 121205, Russia}
\affiliation{Moscow Institute of Physics and Technology, 141700 Dolgoprudny, Russia}

\author{D.~A.~Chermoshentsev}

\author{I.~A.~Bilenko}

\author{F.~Ya.~Khalili}
\email{farit.khalili@gmail.com}

\affiliation{Russian Quantum Center, Skolkovo IC, Bolshoy Bulvar 30, bld.\ 1, Moscow, 121205, Russia}

\date{\today}

\begin{abstract}

We discuss a new class of pure non-Gaussian quantum states of light characterized by trigonal symmetry on the phase plane. We propose the term ``trigonal states'' for them and show that they can be generated using a nondegenerate four-wave mixing process followed by a heralding photon-number measurement of the photon number in one of the two signal modes.

\end{abstract}

\maketitle


\paragraph{Introduction.}

Pure non-Gaussian quantum states of continuous-variable systems \cite{Braunstein_RMP_77_513_2005} have Wigner quasiprobability distributions with negative regions \cite{Wigner_PR_40_749_1932, Hudson_RMP_6_249_1974, Schleich2001}. They are of significant interest for various fields of quantum physics and technologies, including fundamental tests of the applicability of quantum physics to macroscopic objects \cite{Marshall_PRL_91_130401_2003, Romero-Isart_NJP_12_033015_2010, 10a1KhDaMiMuYaCh}, quantum interferometry \cite{Gorshenin_LPL_21_065201_2024, Gorshenin_JOSAB_42_425_2025}, and quantum information processing and communications \cite{Lvovsky_2006_16985}.

Unfortunately, non-Gaussian quantum states are hard to prepare, in particular, because dissipation readily degrades their coherence and non-Gaussian features \cite{Zurek_RMP_75_715_2003}. Fock states and Schr\"odinger cat states are prominent examples of experimentally prepared non-Gaussian states \cite{Ansari_OptExpress_26_2764_2018, Takase_PRA_103_013710_2021, Lio_AQT_5_2200006_2022, Etesse_PRL_114_193602_2015, Huang_PRL_115_023602_2015}.

The optical bistability that arises during spontaneous degenerate parametric generation, can be considered as an analog of the Schr\"odinger cat generation process. Really, within the framework of the classical treatment, small deviations in the initial condition inflate, leading to quasi-random bifurcation of the phase plane trajectory. Recently it was demonstrated experimentally in Ref.\,\cite{Danilin_2601_07378} that in more sophisticated setups with non-degenerate (two-component) pumping, multistable solutions with the axial symmetry of order $k>2$ also exist. In the work \cite{Nougmanov_2606_12749}, this problem was analyzed using the group theory approach.

\begin{figure}
  \includegraphics[width=0.4\textwidth]{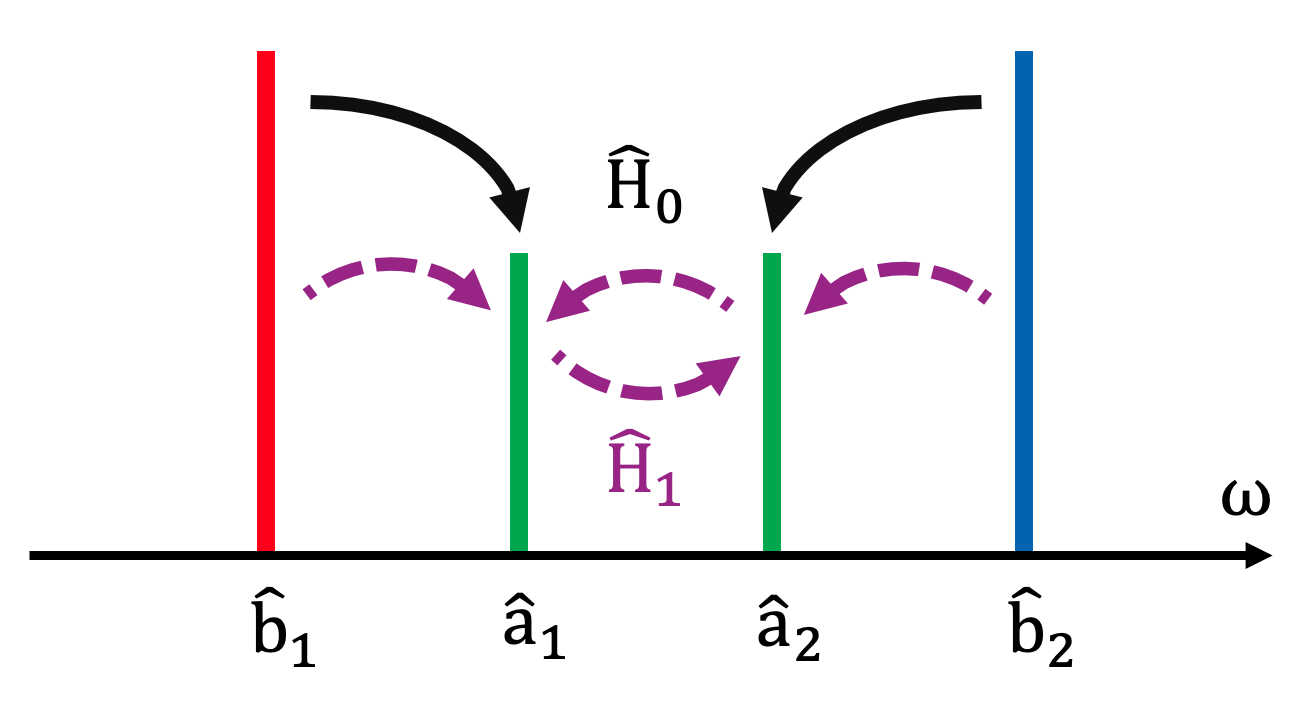}
  \caption{The mode structure of the non-degenerate four-wave mixing system. $\hat{b}_{1,2}$: the pump modes, $\hat{a}_{1,2}$: the signal modes. The mode frequencies are equally spaced.}\label{fig:modes}
\end{figure}

In the quantum case, all branches of evolution in these multistable systems coexist, creating a kind of ``mutlihead cat states''. In Ref.\,\cite{Singh_2606_19085}, a quantum analysis of a four-mode system similar to the one shown in Fig.\,\ref{fig:modes} was done, showing that the resulting quantum state indeed demonstrates the predicted in Ref.\,\cite{Nougmanov_2606_12749} trigonal symmetry. Unfortunately, the numerical calculations were used in that work, and the results were presented as just a set of plots. What is more important, in order to eliminate the unnecessary degrees of freedom, the authors traced them out, which yielded highly mixed states that are much less interesting for applications than pure states.

In general, the $k$-gonal symmetry of a quantum state  means that this state is invariant (up to a constant phase shift) with respect to action of the evolution operator $\exp\bigl(\tfrac{2\pi i}{k}\hat{N}\bigr)$, where $\hat{N}$ is the photon number operator and $k\in\mathbb{N}$. In particular, the Wigner function of such a state possesses the corresponding symmetry on the phase plane. It is easy to show that in the pure state case, the corresponding wave function must have the following form in Fock reprenetation:
\begin{equation}\label{k-gonal}
  \ket{\psi} = \sum_{n=0}^\infty\psi_n\ket{kn+n_0} \,,
\end{equation}
where $n_0$ is a non-negative number. In the mixed state case, it evidently has to be equal to a mix of states of the form \eqref{k-gonal}.

In this paper, in order to reveal the internal structure of the trigonal states, we analytically calculate the non-linear evolution of the non-degenerate four-wave mixing system with two classical pump modes and two quantum signal ones, shown in Fig.\,\,\ref{fig:modes}. Using the perturbative approach, we obtain the aprroximate explicit form of the resulting joint wave function of the two signal modes. We consider the heralding measurement of the photon number in the second mode and show that the resulting pure quantum state of the first mode possesses the trigonal symmetry. We estimate the  parameters of the system that provide well-pronounced trigonal non-Gaussian states and, at the same time, can be in principle implemented in real-world experiments. In order to justify our approximation, we perform also the corresponding exact numerical calculation.

\paragraph{The system.} \label{sec:system-definition}

Consider the four-mode optical system shown in Fig.\,\ref{fig:modes}. Here $\hat{a}_{1,2}$ and $\hat{b}_{1,2}$ are the annihilation operators of the signal and the pump modes. We assume that the eigen frequencies of all four modes are equally spaced. We assume also that the pump modes can be considered as classical ones:
\begin{equation}\label{class_pump}
  \hat{b}_{1,2} \to \beta e^{i\phi_{1,2}} \,,
\end{equation}
where $\beta=\sqrt{N}\gg1$ is the amplitude and $\phi_{1,2}$ are the phases of these modes. This assumption is fully justified by the fact that the nonlinearities of the low-absorbing optical media are weak and therefore strong pumping is required to create noticeable nonlinear effects.

The full Hamiltonian of this system is quite involved. Its general form can be found \eg in Ref.\,\cite{Chembo_PRA_93_033820_2016}. However, the assumption \eqref{class_pump} allows us to simplify it significantly. First, we neglect the self-phase modulation (SPM) and cross-phase modulation (XPM) effects imposed by the pump modes, because they, being classical, can be taken into account in some way. Second, we neglect SPM and XPM effects imposed by the signal modes, because they are small ($\propto\beta^0$). In the rotating wave picture, the remaining Hamiltonian has the following form:
\begin{equation}\label{H}
  \hat{H} = \hat{H}_0 + \hat{H}_1 \,,
\end{equation}
where
\begin{subequations}\label{H_phi}
  \begin{equation}\label{H_00}
    \hat{H}_0 = \hbar\Gamma\hat{a}_1^\dag\hat{a}_2^\dag e^{i(\phi_1 + \phi_2)} + \hc
  \end{equation}
  is the two-mode-squeezing Hamiltonian and the term
  \begin{equation}
    \hat{H}_1 = \hbar\kappa
      (\hat{a}_1^\dag{}^2\hat{a}_2e^{i\phi_1} + \hat{a}_2^\dag{}^2\hat{a}_1e^{i\phi_2})
      + \hc
  \end{equation}
\end{subequations}
originates from the degenerate parametric interactions of the modes $\hat{b}_1$, $\hat{a}_1$, $\hat{a}_2$ and $\hat{b}_2$, $\hat{a}_2$, $\hat{a}_1$, respectively. Note that
\begin{equation}\label{GammaKappa}
  \Gamma = \beta^2\gamma \,, \quad \kappa = \beta\gamma \,,
\end{equation}
where $\gamma$ is the Kerr nonlinearity factor of the media. Therefore,
\begin{equation}
  \frac{\kappa}{\Gamma} = \frac{1}{\beta} \ll 1 \,.
\end{equation}

Eqs.\,\eqref{H_phi} can be further simplified by replacing
\begin{equation}
  \hat{a}_1 := \hat{a}_1\exp\biggl(i\frac{2\phi_1 + \phi_2}{3}\biggr) \,, \quad
  \hat{a}_2 := \hat{a}_2\exp\biggl(i\frac{2\phi_2 + \phi_1}{3}\biggr) \,.
\end{equation}
As a result, we obtain that
\begin{subequations}\label{H01}
  \begin{gather}
    \hat{H}_0 = \hbar\Gamma\hat{a}_1^\dag\hat{a}_2^\dag + \hc \,, \\
    \hat{H}_1 = \hbar\kappa(\hat{a}_1^\dag{}^2\hat{a}_2 + \hat{a}_2^\dag{}^2\hat{a}_1)
      + \hc \label{H_1}
  \end{gather}
\end{subequations}

\paragraph{Evolution of the system.}\label{sec:evolution}

Consider the equation of motion for the evolution operator $\hat{U}(t)$ of our system:
\begin{equation}\label{dU}
  i\hbar\fulld{\hat{U}(t)}{t} = \hat{H}\hat{U}(t) \,.
\end{equation}
Following the standard interaction picture approach, see \eg \cite{Louisell}, we present the evolution operator as follows:
\begin{equation}
  \hat{U}(t) = \hat{U}_0(t)\hat{U}_1(t) \,,
\end{equation}
where the unperturbed part $\hat{U}_0$ and the perturbation $\hat{U}_1$ satisfy the following equations of motion:
\begin{subequations}
  \begin{gather}
    i\hbar\fulld{\hat{U}_0(t)}{t} = \hat{H}_0\hat{U}_0(t) \,,  \label{dU0} \\
    i\hbar\fulld{\hat{U}_1(t)}{t} = \hat{H}_1(t)\hat{U}_1(t) \,, \label{dU1}
  \end{gather}
\end{subequations}
where
\begin{equation}\label{H_1_t}
  \hat{H}_1(t) = \hat{U}_0^\dag(t)\hat{H}_1\hat{U}_0(t) \,.
\end{equation}
is the perturbation Hamiltonian \eqref{H_1} in the interaction picture.

It follows from Eqs.\,\eqref{H_00}, \eqref{dU0} that
\begin{equation}
  \hat{U}_0(t) = \hat{S}(\Gamma t)
\end{equation}
where (for any squeeze factor $r$)
\begin{equation}
  \hat{S}(r) = e^{-ir(\hat{a}_1^\dag\hat{a}_2^\dag + \hat{a}_1\hat{a}_2)}
\end{equation}
is the two-mode squeezing operator. The solution to Eq.\,\eqref{dU1}, in the linear in $\hat{H}_1$ approximation, is the following:
\begin{equation}\label{U_1_approx}
  \hat{U}_1(t) = \hat{I} + \frac{1}{i\hbar}\int_0^t\hat{H}_1(t')\,dt'
\end{equation}
where $\hat{I}$ is the identity operator.

We suppose that initially the system was prepared in the two-mode squeezed state with the initial squeeze factor $r_0$:
\begin{equation}\label{Psi_0}
  \ket{\Psi_0} = \ket{S_2(r_0)} \,,
\end{equation}
where (for any $r$)
\begin{gather}
  \ket{S_2(r)} = \hat{S}(r)\ket{0}_1\otimes\ket{0}_2
  = \frac{1}{\cosh r}\sum_{n=0}^\infty g^n(r)\ket{n}_1\otimes\ket{n}_2 \,, \\
  g(r) = -i\tanh r \,,
\end{gather}
and $\ket{n}_{1,2}$ are the Fock states of the respective modes. This state can be prepared by using the same Hamiltonian \eqref{H} and a short strong pump pulse with the  duration $t_{\rm prep}$ and amplitude $\beta_{\rm prep}$ satisfying the condition $\beta_{\rm prep}t_{\rm prep}\to0$, but $\beta_{\rm prep}^2t_{\rm prep}\to\const$.

The final quantum state is equal to
\begin{equation}
  \ket{\Psi_f} = \hat{U}(t)\ket{\Psi_0} \,.
\end{equation}
Its explicit form, assuming the linear approximation \eqref{U_1_approx}, is calculated in the End Matter section. It can be presented as follows:
\begin{equation}\label{Psi_f}
  \ket{\Psi_f} = \ket{S_2(r_f)} + \ket{\delta\Psi} \,,
\end{equation}
where
\begin{equation}\label{delta_Psi}
  \ket{\delta\Psi} = \frac{K(r_0,r_f)}{\beta\ch^4r_f}
    \sum_{n=0}^\infty R_ng^n(r_f)(\ket{n+3}_1\ket{n}_2 + \ket{n}_1\ket{n+3}_2)
\end{equation}
is the non-linear term,
\begin{multline}\label{K}
  K(r_0,r_f) = \frac{1}{3}\bigl\{\ch^3(r_0) - \ch^3(r_f) \\
    + i[\sh^3(r_f) - \sh^3(r_0)]\bigr\}
\end{multline}
is the effective non-linear interaction factor,
\begin{equation}\label{rf0}
  r_f = r_0 + \Gamma t
\end{equation}
is the final squeeze factor, $t$ is the interaction time, and
\begin{equation}
  R_n = \begin{cases}
          \sqrt{(n+1)(n+2)(n+3)} \,, & n \ge 0 \,, \\
          0 & n < 0 \,.
        \end{cases}
\end{equation}

\paragraph{Heralded state.}\label{sec:heralding}

The two-mode state \eqref{Psi_f} possesses a hidden trigonal symmetry. Really, it can be seen that rotating the two modes in opposite directions by $2\pi/3$, which corresponds to applying the evolution operator $\exp\bigl[\tfrac{2\pi i}{3}(\hat{n}_1 - \hat{n}_2)\bigr]$, leaves the state unchanged. In order to explicitly reveal this symmetry, we consider the heralded quantum state of the first signal mode obtained by measuring the photon number of the second one.

Suppose that the result ``$m$'' is obtained. It is easy to show that the corresponding unnormalized state of the first mode is be equal to
\begin{equation}\label{m_Psi}
  {}_2\bracket{m}{\Psi} = A_m\ket{\psi_{\rm out}} \,,
\end{equation}
where
\begin{equation}\label{psi_out}
  \ket{\psi_{\rm out}}
  = \ket{m}_1 + \psi_{+3}\ket{m+3}_1 + \psi_{-3}\ket{m-3}_1  \,,
\end{equation}
is the corresponding normalized wave function,
\begin{equation}\label{A_m}
  A_m = \frac{g^m(r_f)}{\ch r_f}
\end{equation}
is the probability amplitude of obtaining the result $m$, and
\begin{subequations}\label{psi_pm}
  \begin{gather}
    \psi_{+3} = \frac{K(r_0,r_f)R_m}{\beta\ch^3r_f} \,, \\
    \psi_{-3} = \frac{K(r_0,r_f)R_{m-3}}{\beta g^3(r_f)\ch^3r_f} \,.
  \end{gather}
\end{subequations}
It is easy to see that this wave function has the form of \eqref{k-gonal} with $k=3$, that is, it has the trigonal symmetry.

\paragraph{Discussion.}

\begin{table}
  \begin{ruledtabular}
    \begin{tabular}{ccccc}
                    & $Q_{\rm load}$ & $t=Q_{\rm load}/\omega_0$ & $\gamma$ & $\gamma t$ \\
        \hline
       ${\rm Ca}{\rm F}_2$ &
           $10^9$ & $8\times10^{-7}\,{\rm s}$ & $0.4\,{\rm s}^{-1}$ & $3.2\times10^{-7}$ \\
       ${\rm Si}_3{\rm N}_4$ &
             $10^7$ & $8\times10^{-9}\,{\rm s}$ & $30\,{\rm s}^{-1}$ & $2.4\times10^{-7}$
    \end{tabular}
  \end{ruledtabular}
  \caption{Parameters, used for the estimates, see Ref.\,\cite{20a1BaKhStMaBi}}\label{tab:params}
\end{table}

The prospects of the experimental implementation of the proposed protocol crucially depend on the achievable values of the Kerr nonlinearity factor $\gamma$ and of the optical losses in the resonator that limit the available interaction time $t$.

Let us consider two possible experimental platforms. The first one is the ultra-high-Q whispering gallery mode microresonator fabricated from  calcium fluoride \cite{07a1SaIlMa}; the second one is a planar silicon nitride ring resonator. The parameters relevant for our consideration are listed in Table\,\ref{tab:params}. Mostly, they are based on the Table 1 of Ref.\,\cite{20a1BaKhStMaBi}, with two adjustments. First, we use more modest ``loaded'' values of the quality factors, instead of the ``intrinsic'' ones provided in Ref.\,\cite{20a1BaKhStMaBi}. Second, for the silicon nitride case, we assume very small but feasible volume of the mode $V=10^{-16}\,{\rm m}$, which translates to the value of $\gamma$ indicated in our table (see Eq.\,(61) of Ref.\,\cite{20a1BaKhStMaBi}). In both cases, we obtain the values of the product $\gamma t$ close to $3\times10^{-7}$.

It follows from Eqs.\,\eqref{K} and \eqref{psi_pm}, that in order to obtain well-pronounced nonlinearity, that is the values of probabilities $p_\pm = |\psi_\pm|^2$ comparable with unity, the initial squeezing should be stronger than the final one, $e^{|r_0|}\gg e^{|r_f|}$. On the other hand, (i) the achievable value of squeezing is limited by optical losses; and (ii) the value of $r_f$ can not be too small, because small values of $r_f$ translate to the small heralding probability $|A_m|^2$, see Eq.\,\eqref{A_m}. Taking these limitations into account, we assume the following values for our estimates:
\begin{subequations}\label{param_values}
  \begin{gather}
    r_0 = -2.5 \ (22\,{\rm dB}) \,, \\
    r_f = -1 \ (8.7\,{\rm dB}) \,, \\
    \gamma t = 3\times10^{-7} \,.
  \end{gather}
  It follows from Eqs.\,\eqref{GammaKappa}, \eqref{rf0}, that in this case, there should be
  \begin{equation}
    N \approx 5\times10^6 \,.
  \end{equation}
\end{subequations}
Note that 18\,dB of intracavity squeezing was demonstrated recently in Ref.\,\cite{Karnik_2605_27607}, and it is possible to hope for further progress in this area.

\begin{figure}
  \includegraphics[width=\columnwidth]{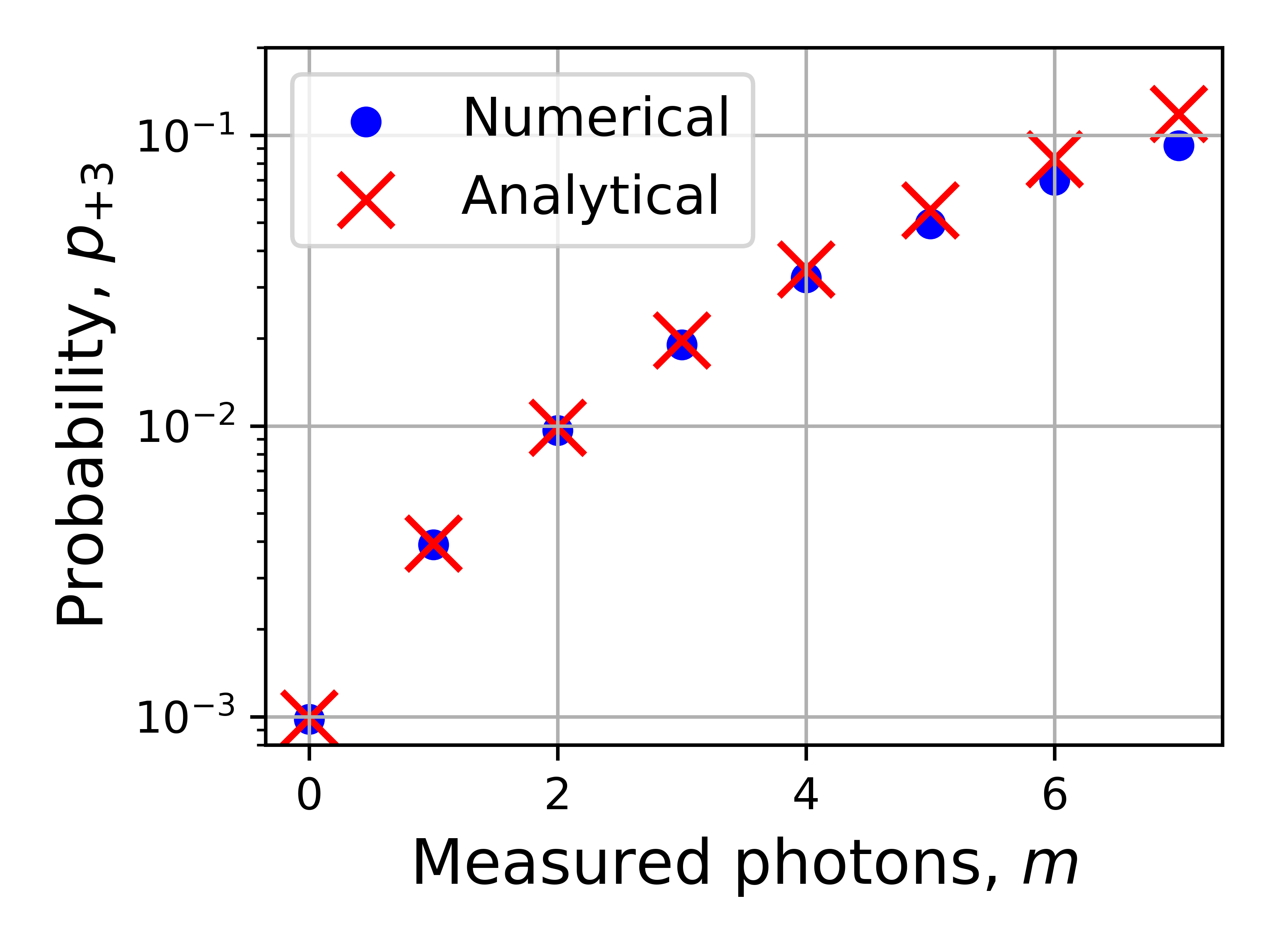}
  \includegraphics[width=\columnwidth]{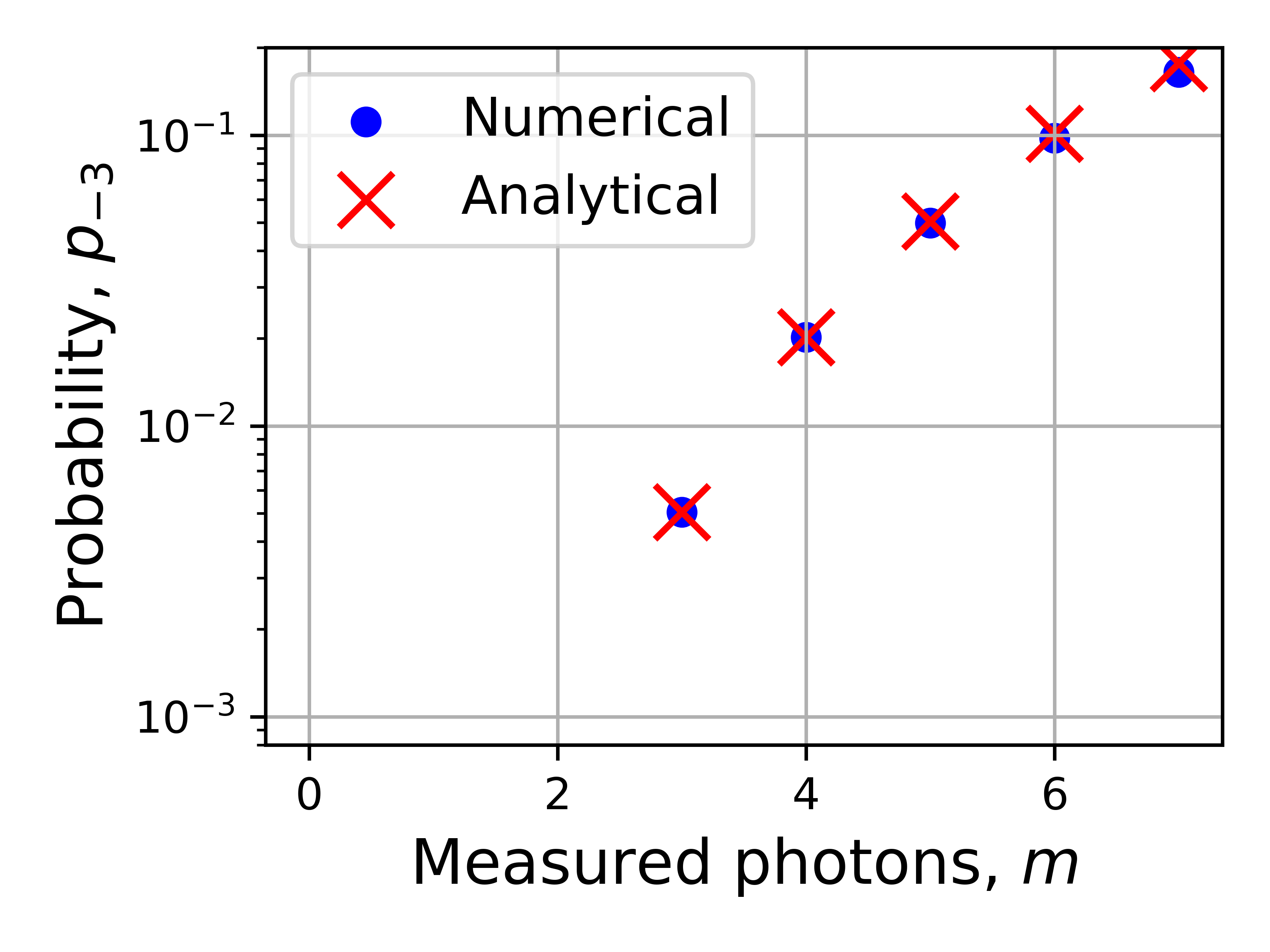}
  \caption{The absolute values the probabilities $p_{+3}=|\psi_{+3}|^2$ (top) and $p_{-3}=|\psi_{-3}|^2$ obtained analytically and numerically as functions of the detected photon number $m$ for the parameters values listed in Eqs.\,\eqref{param_values}.}\label{fig:psi_pm}
\end{figure}

\begin{figure}
  \includegraphics[width=0.4\textwidth]{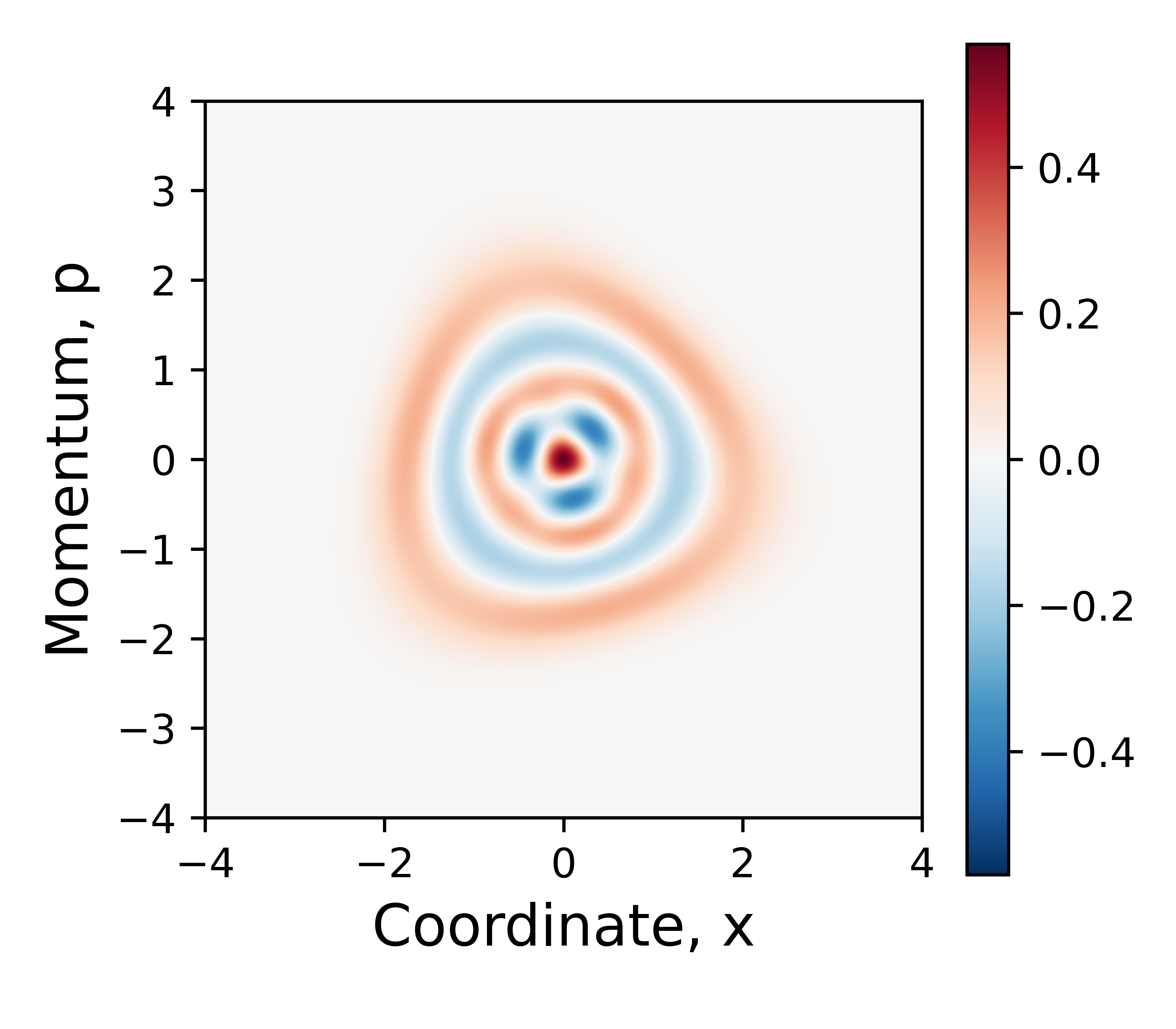}
  \caption{Wigner functions of the heralded output state for $m=4$ and the parameters listed in Eqs.\,\eqref{param_values}.}
  \label{fig:Wigner1}
\end{figure}
\begin{figure}
	\centering
  \includegraphics[width=0.4\textwidth]{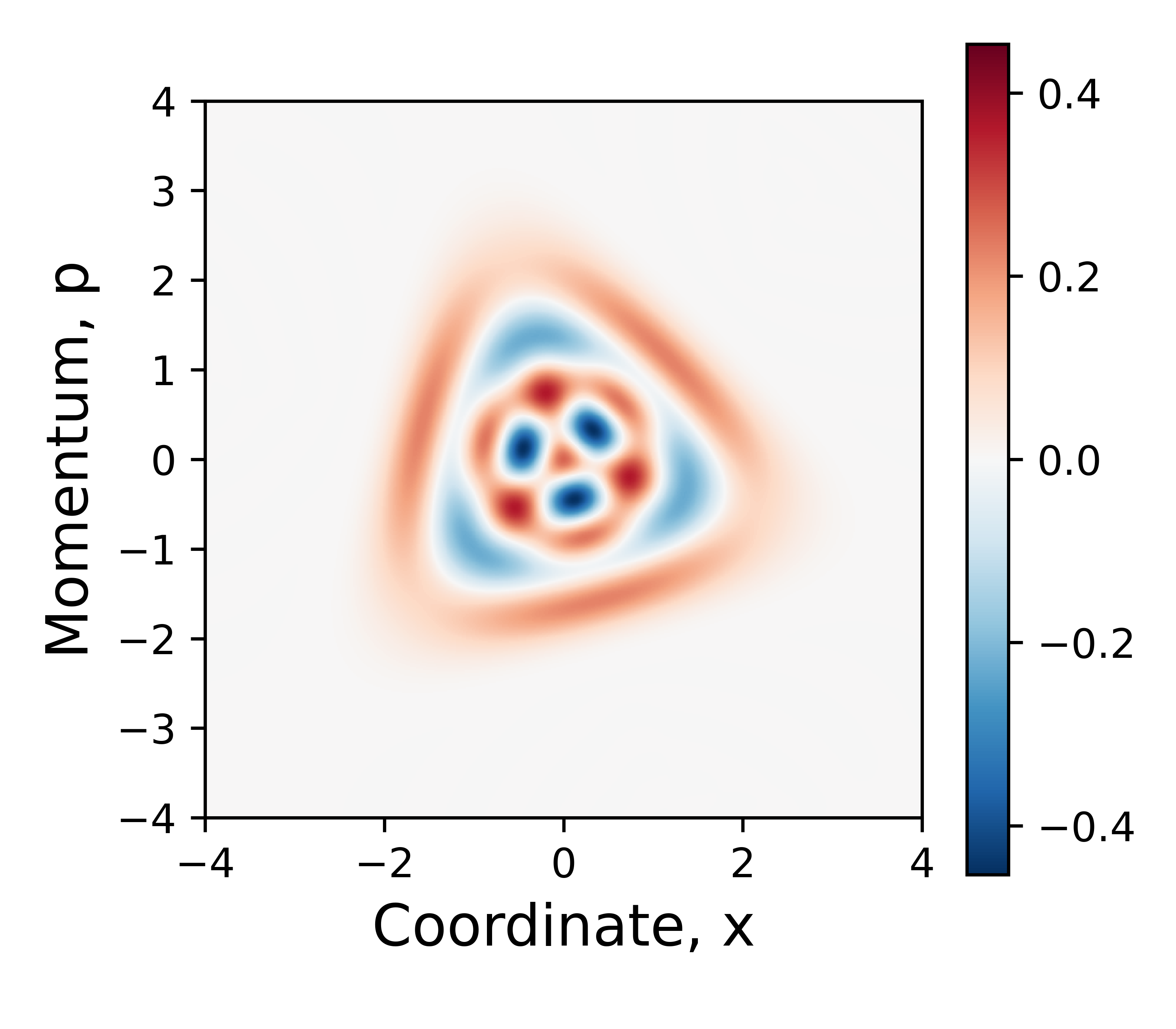}
  \caption{Wigner functions of the heralded output state for $m=4$, $\beta^2=6\times10^6$ and $r_0$, $\gamma\tau$ the same as in Fig.\,\ref{fig:Wigner1}.}
  \label{fig:Wigner2}
\end{figure}

The resulting values of the probabilities $p_\pm$  are plotted in Fig.\,\ref{fig:psi_pm} for the reasonable values of detected photon number $m$. It can be seen that the values comparable with unity could be obtained using the resonators with the demanding but realistic parameters of the setup.

To justify our analytical model and to estimate its applicability limits, we also solve the equation of motion \eqref{dU} and calculate the heralded state \eqref{psi_out} numerically using the {\sf QuTiP} software library \cite{Johansson_CPC_183_8_2012, Lambert_PRep_1153_0370_2026}. The resulting values of the probabilities are also presented in Fig.\,\ref{fig:psi_pm}, showing the good agreement between the both approaches. A noticeable (but still not crucial) discrepancy can be observed only for the case of $m=7$. Evidently, it arises because the approximation \eqref{U_1_approx} is not well-fulfilled for this case.

In Fig.\,\ref{fig:Wigner1}, the corresponding numerically calculated Wigner function for the particular case of $m=4$ is plotted.

Increasing the photon number in the pump modes (and deviating thus form the approximation \eqref{U_1_approx}), more pronounced shape of the Wigner function can be obtained. In particular, in Fig.\,\ref{fig:Wigner2} the numerically calculated Wigner function for the case of $\beta^2=6\times10^6$, with and $r_0$ and $\gamma\tau$ being the same as in Fig.\,\ref{fig:Wigner1}, is plotted.

The success probabilities $|A_m|^2$ in the cases shown in Figs.\,\ref{fig:Wigner1}, \ref{fig:Wigner2} are equal to $\approx4.8\%$ and $\approx1.1\%$, respectively.

It can be seen from these results that the non-Gaussian trigonal quantum states could be prepared using optical resonators with the demanding but realistic parameters.

\acknowledgments
This work was supported by the Russian Science Foundation (Project No. 25-12-00263)


%

\clearpage

\appendix

\section{End Matter}

It follows form Eqs.\,\eqref{H_1_t}, \eqref{Psi_0} that
\begin{multline}
  \hat{H}_1(t')\ket{\Psi_0}
  = \hat{U}_0^\dag(t')\hat{H}_1\hat{U}_0(t')\hat{S}(r_0)\ket{0_2} \\
  = \hat{S}(r_0)\hat{S}^\dag(r_0)\hat{U}_0^\dag(t')\hat{H}_1\hat{U}_0(t')\hat{S}(r_0)
      \ket{0_2}  \\
  = \hat{S}(r_0)\hat{S}^\dag(r')\hat{H}_1\hat{S}(r')\ket{0_2} \,,
\end{multline}
where
\begin{gather}
  \ket{0_2} = \ket{0}_1\otimes\ket{0}_2 \,, \\
  r' = r_0 + \Gamma t' \,.
\end{gather}
Note that $\forall r$,
\begin{subequations}\label{a_12}
  \begin{gather}
      \hat{S}^\dag(r)\hat{a}_1\hat{S}(r) = \hat{a}_1\ch r - i\hat{a}_2^\dag\sh r \,,\\
      \hat{S}^\dag(r)\hat{a}_2\hat{S}(r) = \hat{a}_2\ch r - i\hat{a}_1^\dag\sh r \,.
  \end{gather}
\end{subequations}
Therefore,
\begin{multline}
  \hat{H}_1(t')\ket{\Psi_0}
  = -\hbar\kappa\hat{S}(r_0)(i\ch^2r'\sh r' + \ch r'\sh^2r') \\
      \times(\hat{a}_1^\dag{}^3 + \hat{a}_2^\dag{}^3)\ket{0_2} \,.
\end{multline}
Using Eq.\,\eqref{U_1_approx}:
\begin{equation}
  \hat{U}_1(t)\ket{\Psi_0}
  = \hat{S}(r_0)\Bigl[
        1 + \frac{1}{\beta}K(r_0,r_f)(\hat{a}_1^\dag{}^3 + \hat{a}_3^\dag{}^3)
      \Bigr]\ket{0_2}  \,.
\end{equation}
Then note that
\begin{equation}
  \hat{U}_0(t)\hat{S}(r_0) = \hat{S}(r_f) \,.
\end{equation}
Therefore, the final two-mode quantum state can be presented as follows:
\begin{multline}\label{Psi}
  \ket{\Psi} = \hat{U}_0(t)\hat{U}_1(t)\ket{\Psi_0} \\
  = \hat{S}(r_f)\Bigl[
        1
        + \frac{1}{\beta}K(r_0,r_f)(\hat{a}_1^\dag{}^3 + \hat{a}_3^\dag{}^3)
            \hat{S}^\dag(r)\hat{S}(r)
      \Bigr]\ket{0_2} \\
  = \ket{S_2(r_f)} + \ket{\delta\Psi}  \,,
\end{multline}
where
\begin{equation}
  \ket{\delta\Psi}
    = \frac{1}{\beta}K(r_0,r_f)
        \hat{S}(r_f)(\hat{a}_1^\dag{}^3 + \hat{a}_3^\dag{}^3)\hat{S}^\dag(r_f)\ket{S_2} \,.
\end{equation}
Finally, reversing Eqs.\,\eqref{a_12},
\begin{subequations}
  \begin{gather}
    \hat{S}_2(r_f)\hat{a}_1\hat{S}_2^\dag(r_f) = \ch r_f[\hat{a}_1 - g(r_f)\hat{a}_2^\dag]
      \,,\\
    \hat{S}_2(r_f)\hat{a}_2\hat{S}_2^\dag(r_f) = \ch r_t[\hat{a}_2 - g(r_f)\hat{a}_1^\dag]
      \,,
  \end{gather}
\end{subequations}
(the backward evolution), we obtain Eq.\,\eqref{delta_Psi}.

\end{document}